%% file: main.tex
\documentclass[12pt]{iopart}
\usepackage[utf8]{inputenc}
\usepackage{amsmath}
\usepackage{iopams}
\usepackage{mwe}
\usepackage{graphicx}
\graphicspath{{Figures/}}
\usepackage{url}
\usepackage{epstopdf}
\usepackage{epsfig}
\usepackage{csquotes}
\usepackage{booktabs}
\usepackage[date=year,sorting=none,isbn=false,doi=false,url=false,backend=biber,style=numeric,maxnames=2]{biblatex}
\usepackage{helvet}
\usepackage{esdiff}
\usepackage[noabbrev]{cleveref}
\usepackage{siunitx}
\usepackage{import}
\usepackage{color}
\usepackage{soul}
\usepackage[version=4]{mhchem}
\usepackage[crop=pdfcrop]{pstool}\EndPreamble

\begin{document}

\title{ASCOT simulations of \SI{14}{\mega\electronvolt} neutron rates in W7-X: effect of magnetic configuration}
\author{J. Kontula$^1$, J. P. Koschinsky$^2$, S. Äkäslompolo$^2$,\\T. Kurki-Suonio$^1$ and the W7-X team$^A$}

\address{$^1$ Aalto University, Department of Applied Physics, Espoo, Finland}
\address{$^2$ Max-Planck-Institut für Plasmaphysik, Greifswald, Germany}
\address{$^A$ For the Wendelstein 7-X Team, see the author list: T. Klinger et al., Nucl. Fusion 59, 112004 (2019). doi.org/10.1088/1741-4326/ab03a7.}
\ead{joona.kontula@aalto.fi}

\input{00_abstract.tex}

\input{01_introduction}

\input{02_physics}

\input{03_numerical}

\input{04_results}

\input{05_conclusions}

\input{06_acknowledgments}

\defbibheading{bibliography}[\refname]{%
  \section*{#1}%
  \markboth{}{}
}
\printbibliography

\end{document}

%% file: 00_abstract.tex
\begin{abstract}

Neutron production rates in fusion devices are determined not only by the kinetic profiles but also the fast ion slowing-down distributions.
In this work, we investigate the effect of magnetic configuration on neutron production rates in future deuterium plasmas in the Wendelstein 7-X (W7-X) stellarator.
The neutral beam injection, beam and triton slowing-down distributions, and the fusion reactivity are simulated with the ASCOT suite of codes.
The results indicate that the magnetic configuration has only a small effect on the production of \mbox{\SI{2.45}{\mega\electronvolt}} neutrons from thermonuclear and beam-target fusion. The \mbox{\SI{14.1}{\mega\electronvolt}} neutron production rates were found to be between \mbox{\SI{1.49e+12}{\per\second}} and \mbox{\SI{1.67e+12}{\per\second}}, which is estimated to be sufficient for a time-resolved detection using a scintillating fiber detector, although only in high-performance discharges.

\end{abstract}

%% file: 01_introduction.tex
\section{Introduction}

Stellarators provide a promising alternative to the conventional tokamak fusion reactor concept.
Their advantages are that they are inherently steady-state devices, lack current-driven instabilities, and have many degrees of freedom for optimizing the magnetic geometry with regards to fast ion confinement, low bootstrap current or other desired quantities.
However, due to lack of axisymmetry they require careful optimization of the magnetic field in order to ensure confinement of trapped particles.
Wendelstein 7-X (W7-X) is an advanced stellarator located at the Max Planck Institute for plasma physics in Greifswald, Germany \mbox{\cite{GriegerModularW7-X}}.
It uses an optimized set of 3D shaped modular coils, as well as planar coils, to achieve a wide variety of magnetic configurations \cite{GeigerConfiguration}.
Demonstrating sufficient confinement of fast ions - such as those produced in different ion heating schemes - is one of the main goals of W7-X because any future stellarator reactor relies on the fusion-born alpha particles to transfer their energy into the plasma.
This is only achieved with a long enough confinement time of the alpha particles.
The ion heating methods available for fast ion generation in W7-X are neutral beam injection (NBI) and ion cyclotron resonance heating (ICRH).
The NBI system was commissioned in 2018 using hydrogen, with injection energies up to \mbox{\SI{55}{\kilo\electronvolt}} and power output of up to \mbox{\SI{3.6}{\mega\watt}} \mbox{\cite{WolfOP12}}. Upgrading the NBI power to up to \mbox{\SI{15.7}{\mega\watt}} is possible \mbox{\cite{RustNBI}}. The ICRH system is being installed for future campaigns with power output of approximately \mbox{\SI{1}{\mega\watt}} \mbox{\cite{OngenaICRH}}.

Non-axisymmetric systems require careful optimization of the magnetic field to ensure that radial magnetic field drifts -- such as the $\nabla \boldsymbol{B}$ drift -- are cancelled for trapped particles as well as passing particles.
Even then, the cancellation is only effective near the magnetic axis \mbox{\cite{DrevlakConfinement}}.
Some of the particles can also scatter to 'ripple trapped' or 'superbanana' orbits between two magnetic coils and quickly drift to the wall.
These confinement issues are particularly severe for beam ions in W7-X because of the near-perpendicular NBI injection angle, which means the ions are born predominantly on trapped orbits.
The situation is made worse if the plasma density profile is flat, as most of the ions are then born near the edge of the plasma where the field optimization is less effective.
A flat density profile is typical for gas-fuelled discharges in W7-X, although neutral beam and pellet injection can result in more peaked profiles \mbox{\cite{WolfOP12}}.

While demonstrating optimized fast ion confinement is one of the key goals of W7-X, measuring the fast ion population remains a challenge.
Many currently available fast ion diagnostics in W7-X -- such as the fast ion loss detector \mbox{\cite{OgawaFILD}} and infrared cameras \mbox{\cite{JakubowskiIR}} -- rely on detecting the fast ions only as they leave the plasma and, therefore, cannot observe the confined fast ion population.
Developing diagnostics that can observe the confined fast ions is thus desirable for reaching the goals of W7-X.

To date, all W7-X operation has been with either helium or hydrogen plasmas to avoid fusion reactions that would result in unwanted neutron production; avoiding neutron production lowers both the complexity of experiments and the amount of safety facilities needed.
Operation with deuterium is nevertheless required for future stellarator reactors in order to achieve DT fusion.
To prepare for this, deuterium plasma operation in W7-X is considered to start in the next operational phase, OP2.

Measuring fusion-born neutrons is possible using e.g. a scintillating fiber detector (SciFI) \cite{WurdenTFTRSCIFI, KoschinskyTriton}.
SciFi detectors have been successfully operated at the TFTR \cite{WurdenTFTRSCIFI}, JT-60U \cite{Nishitani_1996}, and LHD \cite{Ogawa_2018} experiments.
The SciFi detector can distinguish between \mbox{\SI{14.1}{\mega\electronvolt}} neutrons from DT reactions and a background consisting of the \mbox{\SI{2.45}{\mega\electronvolt}} neutrons from DD reaction and gamma rays, and therefore it can measure the neutron flux from DT reactions exclusively. 
If the flux is sufficiently high, even time-resolved measurements can be made.
When cross-calibrated with either the overall neutron rate or the shot-integrated \SI{14.1}{\mega\electronvolt} neutron rate, the DT fusion rate in the plasma can be calculated from the SciFi measurement.
The overall neutron rate is commonly monitored with fission chambers in the frame of radiation safety \cite{SchneiderNeutronDiagnostics, Wiegel2013NeutronMonitoring}, while the shot-integrated neutron rate can be obtained from neutron activation systems.

The DT fusion rate in the plasma is directly related to the triton slowing-down distribution.
This makes it conceptually possible to use the SciFi detector as a fast ion diagnostic for triton confinement studies.
A key quantity is the triton burn-up ratio, i.e., the ratio of DD fusion born tritons that subsequently undergo DT fusion.
Earlier simulations have shown the feasibility of using SciFi for fast ion diagnostic purposes in W7-X, but have neglected explicit particle losses and configuration effects \cite{KoschinskyTriton}.

The magnetic configuration influences the distribution of beam ions via orbit losses, particularly losses of trapped ions \cite{AkaslompoloW7-XWallLoads}.
Since the beam-target fusion rates are directly proportional to the beam ion slowing-down distribution, the magnetic configuration could also have an effect on the DD -- and indirectly the DT -- fusion rates.
This could potentially limit the neutron flux to the detector and making time-resolved measurements impossible.
On the other hand, if the neutron flux differences between configurations are large enough to be measured using the SciFi detector, this could provide insight into fast ion confinement properties in different configurations.
In any case, validating the SciFi detector necessitates accurate modeling of both the beam deuterium and triton birth profiles and slowing-down distributions.

In this work, we investigate the effect of the magnetic configuration on neutron production rates in W7-X.
NBI ion and triton confinement as well as neutron production are simulated in order to assess the relative \SI{2.45}{\mega\electronvolt} and \SI{14.1}{\mega\electronvolt} neutron rates and, subsequently, the triton burn-up ratio.
In \cref{sec:physics}, the physics and processes of neutron generation and measurement in W7-X are discussed.
The numerical schemes and simulation tools are introduced in \cref{sec:numerical}. The results are compiled in \cref{sec:results}, where the results of NBI injection, fast ion slowing-down, and fusion rate simulations are shown. Additionally, the sensitivity to kinetic profiles -- which are expected to dominate over configuration effects -- is briefly addressed in \cref{sec:results:kinetic}. The findings are summarized in \cref{sec:conclusions}, which also includes suggestions for further work on this topic.

%% file: 02_physics.tex
\section{Physics of neutron generation in W7-X}
\label{sec:physics}

There are two reaction channels for pure DD fusion, each of which has approximately equal probability:
\begin{align*}
{}^{2}_{1}\mathrm{D} + {}^{2}_{1}\mathrm{D} &\;\longrightarrow\; {}^{1}_{1}\mathrm{p} (\SI{3.02}{\mega\electronvolt}) + ^{3}_{1}\mathrm{T} (\SI{1.01}{\mega\electronvolt}) & (\SI{50}{\%})\\
    {}^{2}_{1}\mathrm{D} + {}^{2}_{1}\mathrm{D} &\;\longrightarrow\; {}^{1}_{0}\mathrm{n} (\SI{2.45}{\mega\electronvolt}) + {}^{3}_{2}\mathrm{He} (\SI{0.82}{\mega\electronvolt}) & (\SI{50}{\%})
\end{align*}
The \SI{1.01}{\mega\electronvolt} tritons born from the first reaction channel can, in turn, partake in DT fusion reactions with deuterium ions:
\begin{align*}
    {}^{2}_{1}\mathrm{D} + {}^{3}_{1}\mathrm{T} &\;\longrightarrow\; {}^{1}_{0}\mathrm{n} (\SI{14.1}{\mega\electronvolt}) + {}^{4}_{2}\alpha (\SI{3.5}{\mega\electronvolt})
\end{align*}

The DD fusion reactions require high reactant energies in order to have significant reaction probabilities.
These energies can be found in the high-energy tail of the thermal ion distribution, or in energetic minority populations.
Currently, the main source of high-energy ions in W7-X is the neutral beam injection (NBI) system.
The NBI system consists of two symmetrically placed NBI boxes, each with 4 positive ion neutral injectors (PINIs).
The sources are notably perpendicular, with the injected pitch angles deviating only \SI{17}{\degree} or \SI{27}{\degree} from fully perpendicular, depending on the PINI.
In the latest W7-X campaing, the NBI system was successfully commissioned for hydrogen injection with two PINIs.
The NBI system in W7-X will be extended for more sources and deuterium operation in future experiments \cite{RustNBI}.
The parameters for deuterium injection are shown in \cref{tab:nbi}.

\begin{table}[ht]
\centering
\caption{Planned parameters for W7-X deuterium neutral beam injection. The species mix denotes the fraction of beam particles born at different divisions of the beam acceleration energy \mbox{$E_b = \SI{60}{\kilo\electronvolt}$.}}
\begin{tabular}{@{}lll@{}}
\toprule
Acceleration voltage        & \SI{60}{\kilo\volt}       \\
Max. number of PINIs        & 8                         \\
Beam power (all sources)    & \SI{15.7}{\mega\watt}     \\
Species mix (particle fraction) & \SI{65}{\%} ($E_b$); 25\% ($E_b/2$); 10\% ($E_b/3$)  \\ \bottomrule
\end{tabular}
\label{tab:nbi}
\end{table}

With deuterium beams, there are three different neutron production channels for DD fusion: thermonuclear, beam-target, and beam-beam (B-B). 
Thermonuclear fusion occurs in the high-energy tail of the thermal ion population, beam-target fusion between energetic beam ion and thermal ion populations, and beam-beam fusion between two beam ions.
For DT fusion, the two main production channels are triton-plasma and triton-beam reactions, representing reactions with thermal and fast deuterium, respectively.
It is also possible to consider DT reactions of thermal (slowed-down) tritons separately, although the reaction probabilities for thermal tritons are much lower than for the fast tritons.

For reactions between two Maxwellian ion distributions at thermal equilibrium, the fusion reactivity is given by the product \cite{BoschCrossSections}
\begin{align}
R_{A-B} &= \frac{n_A n_B}{1 + \delta_{AB}} \langle\overline{\sigma v}\rangle_{AB}, \\
\langle\overline{\sigma v}\rangle_{AB} &\propto T^{-2/3} e^{T^{-1/3}},
\end{align}
where the ion species temperatures are $T_A = T_B = T$ and the delta function ensures that reactions from the same population are not doubly counted.
The fusion rate coefficient $\langle\overline{\sigma v}\rangle_{AB}$ has a strong temperature dependency.
For non-Maxwellian distributions, the cross sections depend similarly on the center-of-mass energy of the reactants.
Due to the relatively low thermal ion energies in W7-X, the beam-target fusion is expected to be the primary production channel, making the beam slowing-down distribution critical in determining the total fusion rate.

By using an energy-discriminating neutron detector, such as a scintillating fiber (SciFi) detector \cite{WurdenTFTRSCIFI, Nishitani_1996} and a common detector monitoring the overall neutron rate, it is possible to separately measure the \SI{2.45}{\mega\electronvolt} and \SI{14.1}{\mega\electronvolt} neutron rates.
The DT fusion rate is directly proportional to the triton density in the plasma, which makes it possible to use SciFi, combined with numerical models of the triton slowing-down distribution, to assess the triton burn-up ratio in the plasma.
The triton burn-up ratio depends on the fusion reactivity, the triton slowing-down time, and possible triton orbit losses.
The first two are determined by the kinetic profiles, while the losses are determined by the magnetic configuration.

The magnetic configuration space studied in this work is spanned by the W7-X reference magnetic configurations. These consist of high mirror (HM), characterized by low neoclassical transport and good fast ion confinement; standard (STD) which is located at the center of the configuration space; low mirror (LM) in which the magnetic field strength is nearly constant along the magnetic axis; inward shifted (IS) and outward shifted (OS) plasma configurations; low iota (LI) and high iota (HI) configurations with correspondingly lower and higher edge rotational transform; and low shear (LS) which has a relatively flat rotational transform profile.
The same configurations have previously been studied for hydrogen beam ion confinement with ASCOT \cite{AkaslompoloW7-XWallLoads}, and as such provide a good reference case for deuterium and triton confinement studies.

%% file: 03_numerical.tex
\section{Numerical approach to triton burnup modeling}
\label{sec:numerical}

\begin{figure}[tb]
\centering
\includegraphics[width=0.6\linewidth]{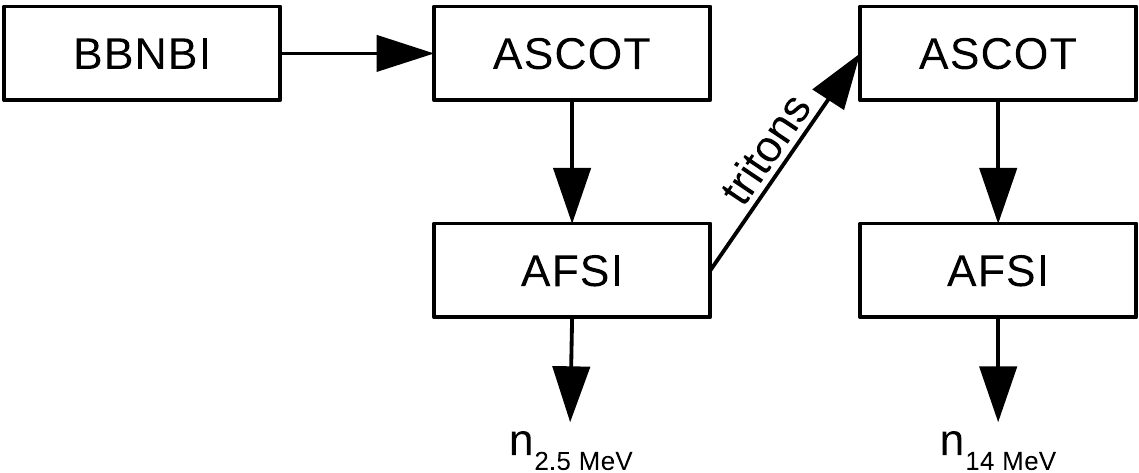}
\caption{Flowchart of the data and tools used. The black arrows represent simulation I/O.}
\label{fig:flowchart}
\end{figure}

For simulating all aspects of triton burnup in W7-X, a staged numerical approach was adopted.
First, the neutral beam injection is simulated with the BBNBI code \cite{AsuntaBBNBI}.
Next, a slowing-down simulation of the ionized NBI population was done using the Monte-Carlo orbit following code ASCOT \cite{VarjeASCOT5}.
ASCOT is a comprehensive tool for fast ion simulations in both tokamaks and stellarators, and has already been successfully used to predict the beam ion power loads on the plasma-facing components in W7-X \cite{AkaslompoloValidating,AkaslompoloEmptyTorus,AkaslompoloArmoring}.
The deuterium slowing-down distribution is used as input for the ASCOT Fusion Source Integrator AFSI \cite{SirenAFSI}, which calculates the DD fusion rates for both reaction channels.
In this work, only the DD $\rightarrow$ pT channel was considered.
The slowing-down distribution of the \SI{1.01}{\mega\electronvolt} triton population is again calculated with ASCOT, and another AFSI simulation performed to get the \SI{14.1}{\mega\electronvolt} neutron birth profiles.
The whole simulation chain is illustrated in \cref{fig:flowchart}.
For Fig. 4, 5, 7 and 9 the values are calculated from distribution functions which are collected between equidistant flux intervals. This means that they are average values, with the data point in the plots located at the center of the interval. 

\begin{figure}[tb]
\centering
\includegraphics[width=0.7\textwidth]{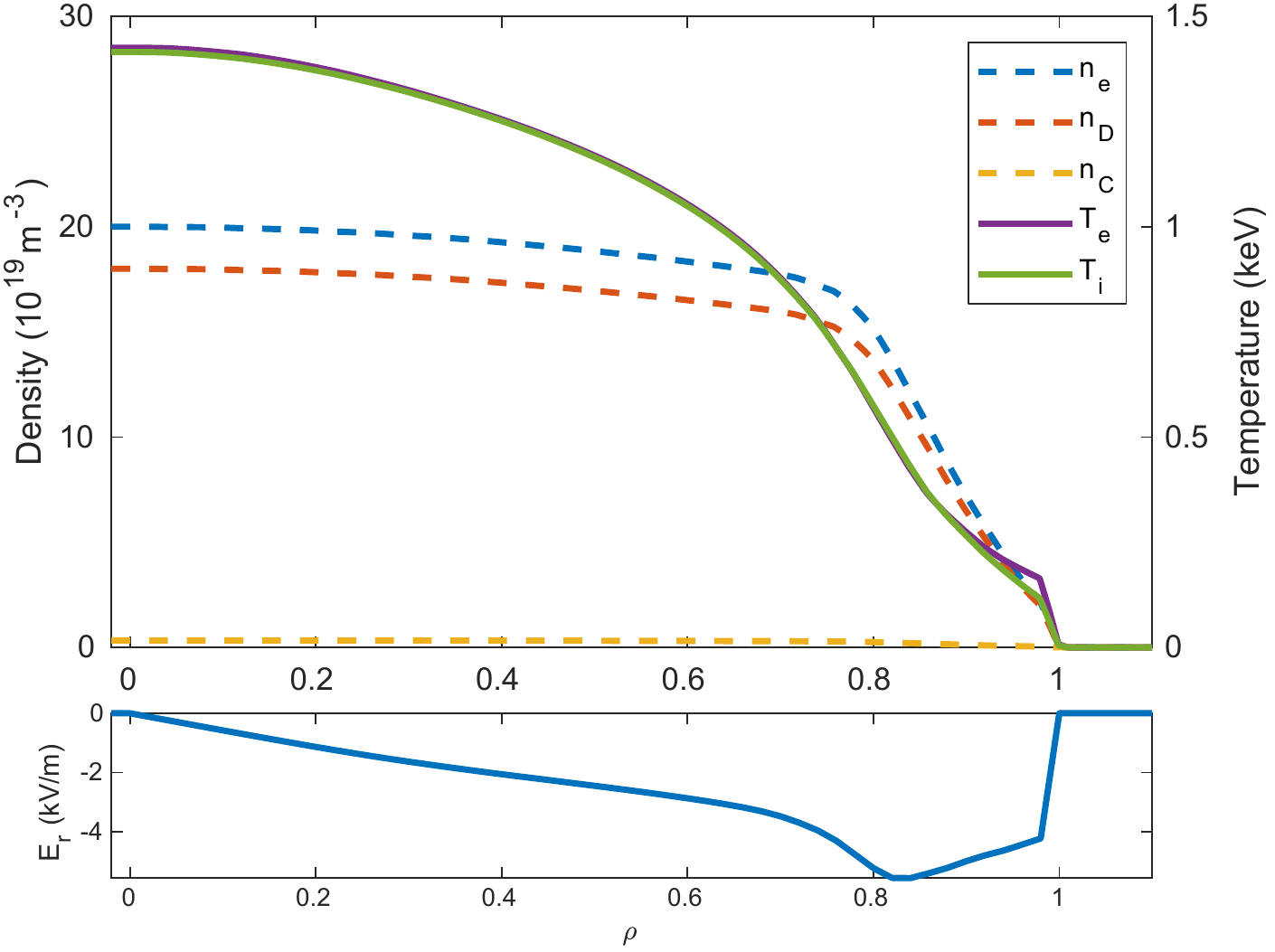}
\caption{Plasma profiles used for the simulations as a function of $\rho$ (square root of normalised toroidal flux). In the upper frame the dashed lines are the species densities; blue for electrons, red for deuterium ions, and yellow for carbon ions. The solid lines show the electron and ion temperatures. The radial electric field, shown in the lower frame, is fixed to zero outside $\rho = 1$ while the plasma profiles are extrapolated linearly from the last two data points.}
\label{fig:profiles}
\end{figure}

The plasma backgrounds -- ion and electron densities and temperatures and radial electric field -- were obtained from simulations using a combination of the NTSS \cite{TurkinNTSS}, DKES \cite{HirshmanDKES}, and TRAVIS \cite{MarushchenkoTRAVIS} codes.
A fixed volume-averaged beta of $\langle\beta\rangle = \SI{2}{\%}$ was used for the plasma profiles, and carbon assumed as the sole impurity.
The radial profiles are shown in \cref{fig:profiles}.
The plasma profiles were originally simulated for a standard configuration hydrogen plasma \cite{DrevlakConfinement}, but for the purposes of the deuterium simulations the primary plasma species was changed to deuterium while keeping the profile shape unchanged.
The same plasma profiles were used for all of the magnetic configurations.
This is not completely rigorous since the magnetic configuration has an effect on the kinetic profiles \cite{DinklageConfiguration}.
However, using the same profiles allows us to isolate the effect of the magnetic configuration.
This effect might otherwise be obscured by the profile changes.

For the deuterium beam slowing-down simulation, the guiding center approximation was used.
For the \SI{1.01}{\mega\electronvolt} triton simulation, the guiding center approximation is not justified since the width of the triton gyro orbits at \SI{2}{\tesla} -- the lowest magnetic field inside the plasma region of W7-X -- can be up to \SI{19}{\centi\meter}, which is of the same order as the plasma minor radius -- approximately \SI{53}{\centi\meter}.
This necessitated a full gyro-orbit simulation for the tritons.

%% file: 04_results.tex
\section{Results}
\label{sec:results}

For the BBNBI beam injection simulations, the full beam power of \SI{1.96}{\mega\watt} for each PINI -- \SI{15.7}{\mega\watt} in total -- was used as the injected power. Beam duct transmission was accounted for by including the beam duct geometry in the wall model.
It should be noted that using all eight available PINIs is an optimistic estimate, since the number of power sources available for the NBI system are shared by ECRH and ICRH heating, and present power resources would be insufficient for simultaneous operation of ECRH heating and all PINIs.
A more realistic estimate of available PINIs would be four, i.e., half the power presented here.
This is also the amount of sources planned for the next W7-X operational phase, OP2.
Nevertheless, using all eight PINIs provides an upper limit for the beam power.

The trajectories of the injected neutrals were simulated until they were ionized or hit the beam dump on the opposite wall.
In the latter case the particle was recorded as a shine-through particle.
The shine-through fraction in all of the cases was nearly identical -- between \SIrange{12.7}{13.0}{\percent} -- as the shine-through fraction is mainly determined by the plasma density profiles.
The only difference in shine-through between configurations is caused by variations in the shape of the plasma. 

\begin{figure}[tb]
\centering
\includegraphics[width=0.6\textwidth]{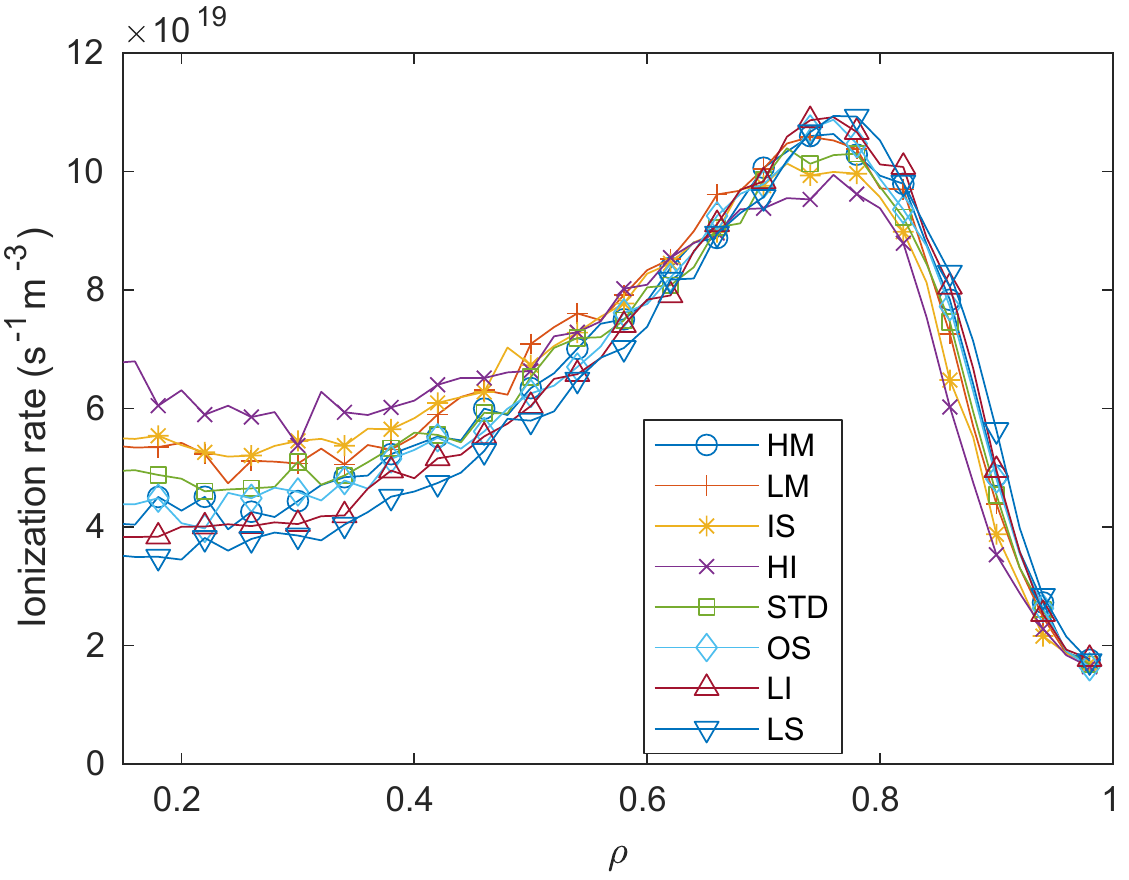}
\caption{Beam ion birth profiles for the different magnetic configurations. The profiles are calculated as histograms from the birth locations of NBI ions. The total number of ionized particles is approximately equal in all of the configurations.}
\label{fig:beam_profiles}
\end{figure}

A limitation of the ASCOT flux coordinate $\rho$ (square root of normalised toroidal flux) used in the simulations is that it is interpolated from a regular 3D grid.
As the flux surfaces are packed more tightly near the plasma core, the spatial resolution of $\rho$ near the magnetic axis is poor.
Due to this limitation, the radial distributions inside $\rho = 0.15$ were averaged and only one value calculated for volume-integrated quantities.
The error made by averaging is small due to the rapidly decreasing plasma volume near the axis.
The plasma density reaches a high-density plateau already at $\rho = 0.8$ as illustrated by \cref{fig:profiles}.
The relatively low-energy beam neutrals get ionized in large numbers already at this radius, leading to birth profile that peaks at the edge. 
The full radial ion birth rate profiles are shown in \cref{fig:beam_profiles}.
Even though the same kinetic profiles were used for all of the configurations, there are still small differences in the birth profiles.
The differences stem from the fact that the shape of the plasma -- and thus the kinetic profiles along the beam sight line -- is different for each configuration.
In addition, the volume between each isosurface of $\rho$, i.e., the shell volumes, are different for each configuration, leading to differences in the ion birth density.
The large variation in the profiles near the axis are mainly caused by the small shell volumes, which exaggerates the differences between configuration.

The observed fast ion birth profile is unfavorable for fast ion confinement.
W7-X is optimized for improved fast ion confinement only near the axis, while the BBNBI simulations predict that over \SI{80}{\%} of the particles are born outside $\rho = 0.5$ (half the minor radius), corresponding as close as \SI{10}{\centi\meter} from the last closed flux surface (LCFS).
In comparison, the particle Larmor radius at this radial location can be up to \SI{2}{\centi\meter} and the banana width up to \SI{15}{\centi\meter}.
Any banana orbits that open outward radially are thus susceptible to hitting the wall; the distance between the LCFS and the divertors is at worst less than \SI{12}{\centi\meter}.

\subsection{DD fusion rates}
\label{sec:results:dd-fusion}

\begin{figure}[tb]
\centering
\includegraphics[width=0.74\textwidth]{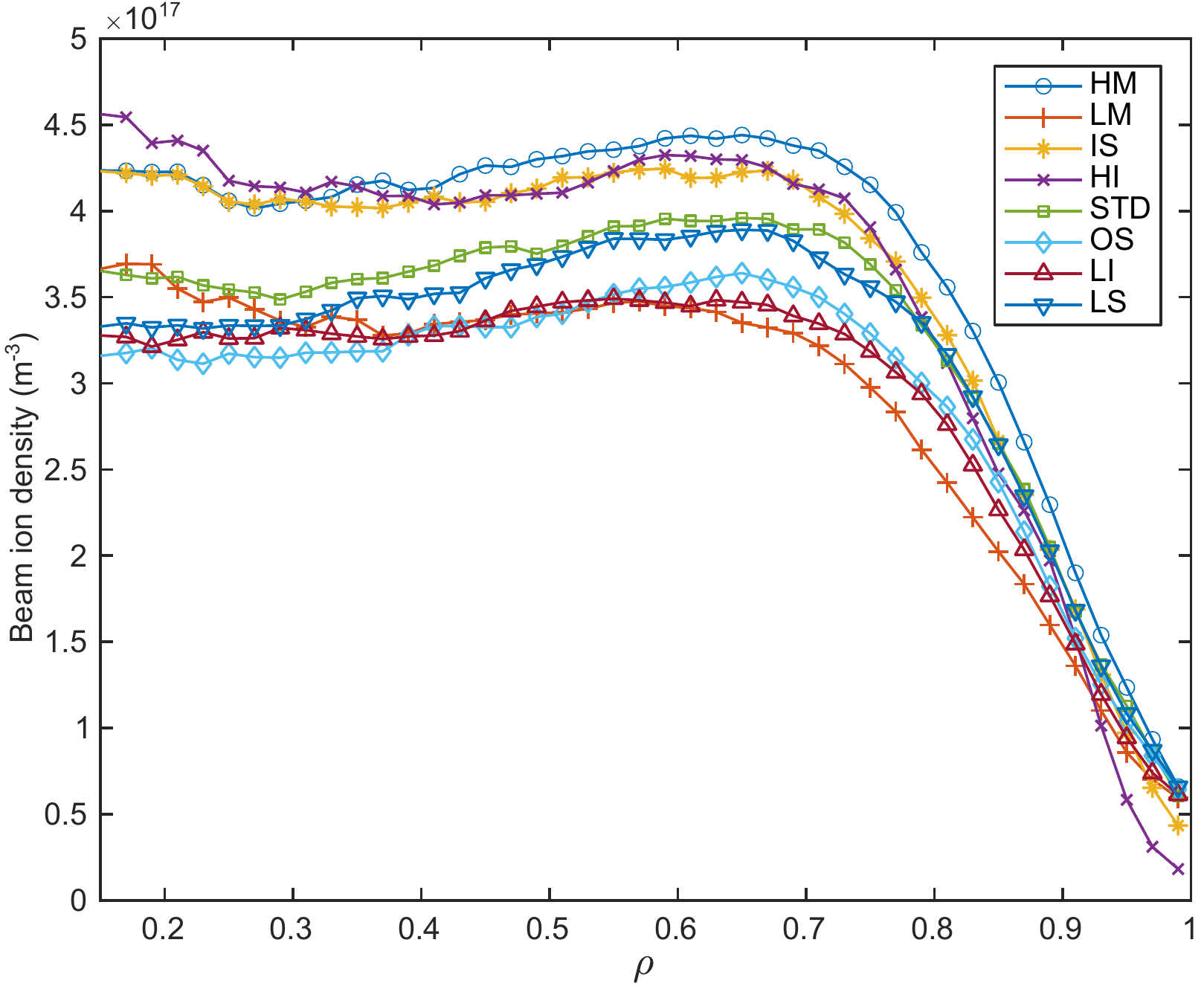}
\caption{Slowing-down distribution of injected deuterium for different magnetic configurations. The peaked deuterium birth profile is flattened by the larger loss fraction of particles near the edge of the plasma.}
\label{fig:beam_dist_rho}
\end{figure}

The deuterium beam slowing-down simulations were done using \num{100000} markers per simulation.
The markers were followed using the guiding-center formalism until they passed the LCFS or their energy was less than twice the local thermal energy, in which case the particle was considered thermalized.
For the particles crossing the LCFS, the simulation was continued in full gyro-orbit mode to account for wall collisions accurately.
The resulting radial beam-ion distributions are shown in \cref{fig:beam_dist_rho}.
Unlike the deuterium birth profiles, the slowing-down density profiles have a flat shape with the density rapidly decreasing outside $\rho = 0.8$, the profile shape being similar between the configurations.
The profile flattening is due to the beam particle loss fraction increasing rapidly towards the edge of the plasma, the loss fraction being 2 to 5 times higher at $\rho = 0.8$ than at the axis.
The slowing-down profiles show differences across the whole plasma even though the kinetic profiles used -- and thus the fast ion slowing-down times -- were identical, suggesting that the differences stem from beam ion orbit properties.

The total beam power loss fraction varies widely between the configurations, from \SI{20.1}{\%} in the high mirror configuration to \SI{39.4}{\%} in the low mirror configuration.
In all of the configurations the losses for deuterium NBI was found to be higher than for hydrogen NBI with the same kinetic profiles \cite{AkaslompoloW7-XWallLoads}.
This is most likely due to the wider orbits of the more massive deuterium ions: the power loss difference in all of the configurations is approximately \SI{40}{\percent}, which is also the difference between the gyroradii of hydrogen and deuterium.
Even though the deuterium simulation was made using the guiding-center formalism, the effect of increasing the particle mass can be seen in the guiding-center collision operator, banana orbit widths, and also the wall collision checks which use the full gyro orbit for accuracy.
It should also be noted that less than \SI{2}{\%} of the particles were lost during their first orbit.
Most of the fast ion losses were via drift motion.

\begin{figure}[tb]
\centering
\includegraphics[width=0.74\textwidth]{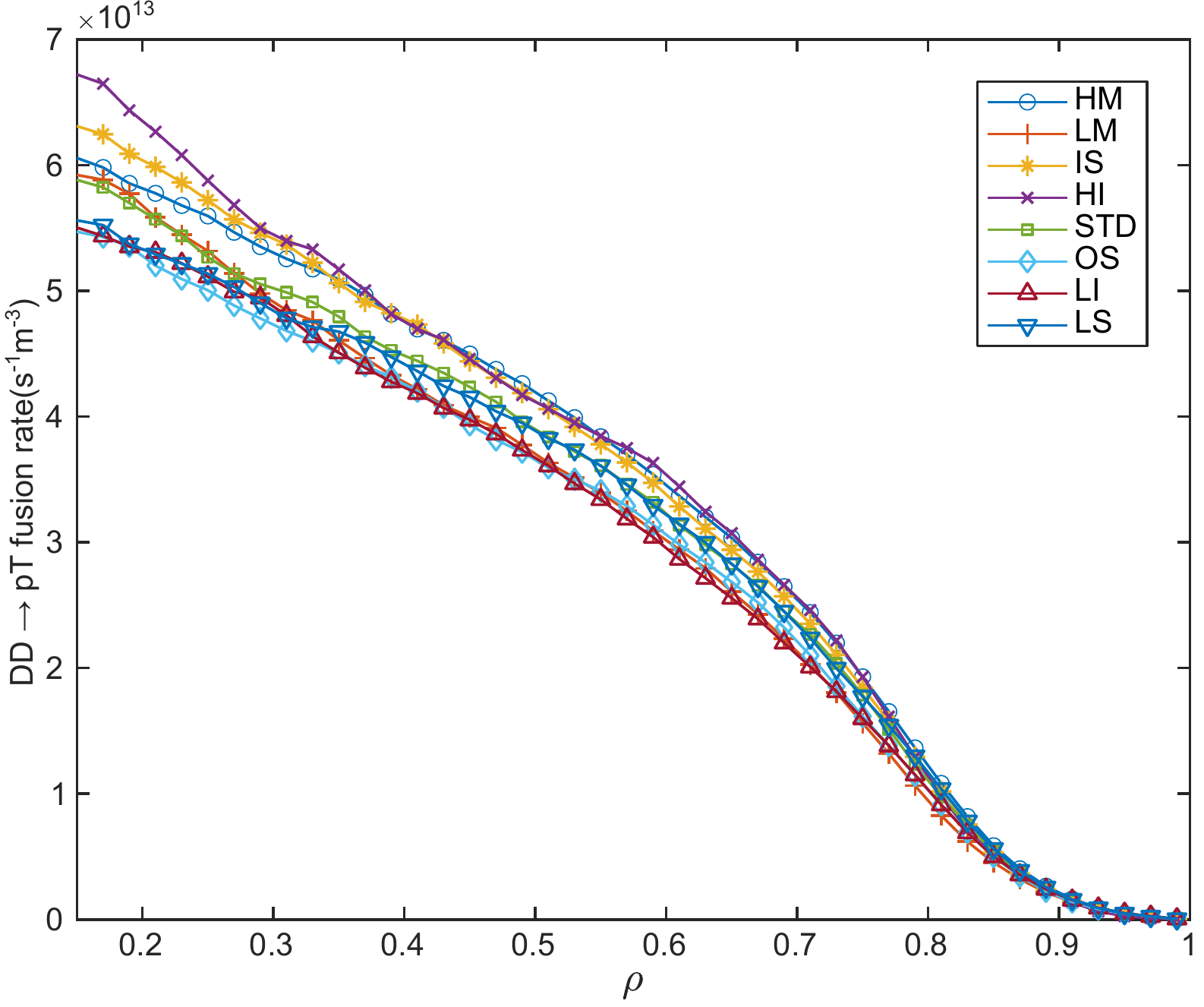}
\caption{Radial DD $\rightarrow$ pT fusion rate or, equivalently, \SI{2.45}{\mega\electronvolt} neutron and \SI{1.01}{\mega\electronvolt} triton birth rate profiles as a function of $\rho$ for the different magnetic configurations. The higher fusion reactivity at higher temperatures causes the profile to peak at the magnetic axis.}
\label{fig:fusion_profiles_dd}
\end{figure}

The fusion rates between the full 5D slowing-down and thermal distributions were calculated with AFSI and converted to 1D $\rho$-profiles for visualization purposes.
Only the DD $\rightarrow$ pT reaction channel was calculated: the rates are identical to the other reaction channel.
The AFSI results indicate that the triton birth distribution is peaked in the center of the plasma, as shown in \cref{fig:fusion_profiles_dd}.
This is due to the fact that both the ion temperature and the beam ion mean energy -- the high energy component of the beams penetrates the plasma more easily -- are centrally peaked and the fusion cross sections have a strong dependence on the ion center-of-mass energy.

\begin{figure}
\centering
\includegraphics[width=0.45\textwidth]{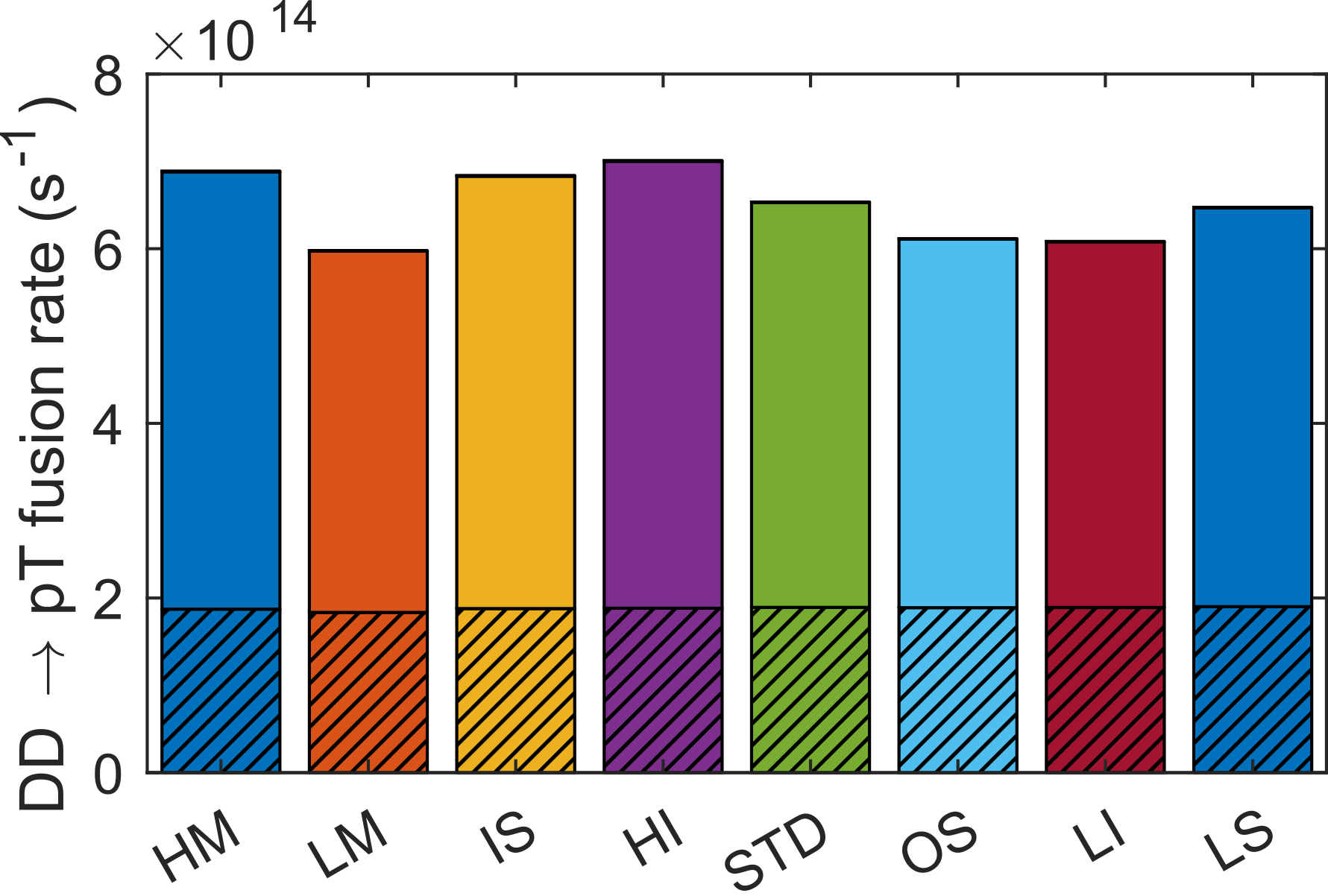}
\caption{Total DD $\rightarrow$ pT fusion rates or, equivalently, \SI{2.45}{\mega\electronvolt} neutron and \SI{1.01}{\mega\electronvolt} triton birth rates in different magnetic configurations. The rates are split according to the production channels: thermonuclear (solid) and beam-target (hash). The beam-beam reactions constitute less than \SI{2}{\percent} of the total fusion rates and are thus practically invisible in the plot.}
\label{fig:reaction_contributions_dd}
\end{figure}

Of all the simulated scenarios, the high-iota configuration had the highest total triton (and \SI{2.45}{\mega\electronvolt} neutron) production rate of \SI{7.01e14}{\per\second}, while the low mirror configuration had the lowest rate at \SI{5.98e14}{\per\second}.
A comparison between all magnetic configurations is shown in \cref{fig:reaction_contributions_dd}, where the production channels are also separated.
In all configurations, the majority of fusion reactions come from the beam-target production channel and the beam-target fusion rate is almost directly proportional to the mean beam ion density in the plasma, which is in turn determined by the beam ion confinement properties of the configurations.
The thermonuclear contribution is virtually identical between the different configurations as the plasma profiles were kept constant.
The beam-beam reactions constitute less than \SI{2}{\%} of the total fusion rate due to the low density of the beam ions.

\subsection{DT fusion rates}
\label{sec:results:dt-fusion}

\begin{figure}[tb]
\centering  
\includegraphics[width=0.75\textwidth]{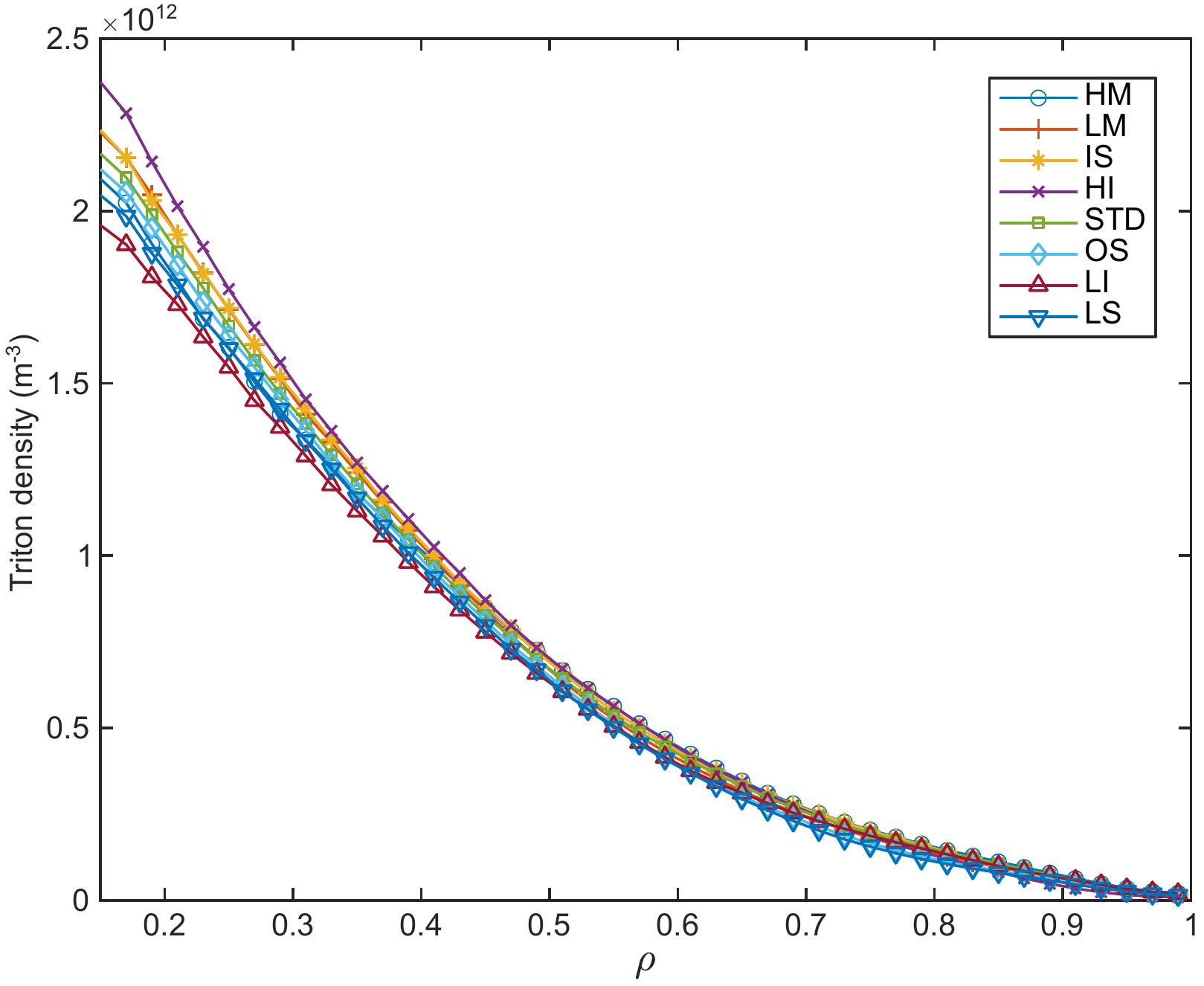}
\caption{Slowing-down distribution of tritons for the different magnetic configurations. The profiles are highly peaked in the plasma core, where the triton birth rate is the highest and loss fraction lowest.}
\label{fig:triton_dist_rho}
\end{figure}

For the triton slowing-down simulation, \num{100000} markers were randomly sampled from the 5D triton birth distribution and simulated with ASCOT using the full-orbit formalism.
The resulting radial triton slowing-down distributions are shown in \cref{fig:triton_dist_rho}.
The total triton power loss fraction was between \SIrange{42}{51}{\%} in the different configuration, which is much higher than for deuterium ions.
This is not surprising, since the triton Larmor radius can be up to half the plasma minor radius.
The loss fraction increases radially from over \SI{50}{\%} for tritons born near the axis to over \SI{80}{\%} for those born outside $\rho = 0.9$.
Due to this and the centrally peaked triton birth profiles, the triton slowing-down profiles are highly peaked in the plasma core.
The profile shape is similar between the configurations, and the differences lie in the total integrated number of tritons in the plasma.

\begin{figure}[tb]
\centering
\includegraphics[width=0.77\textwidth]{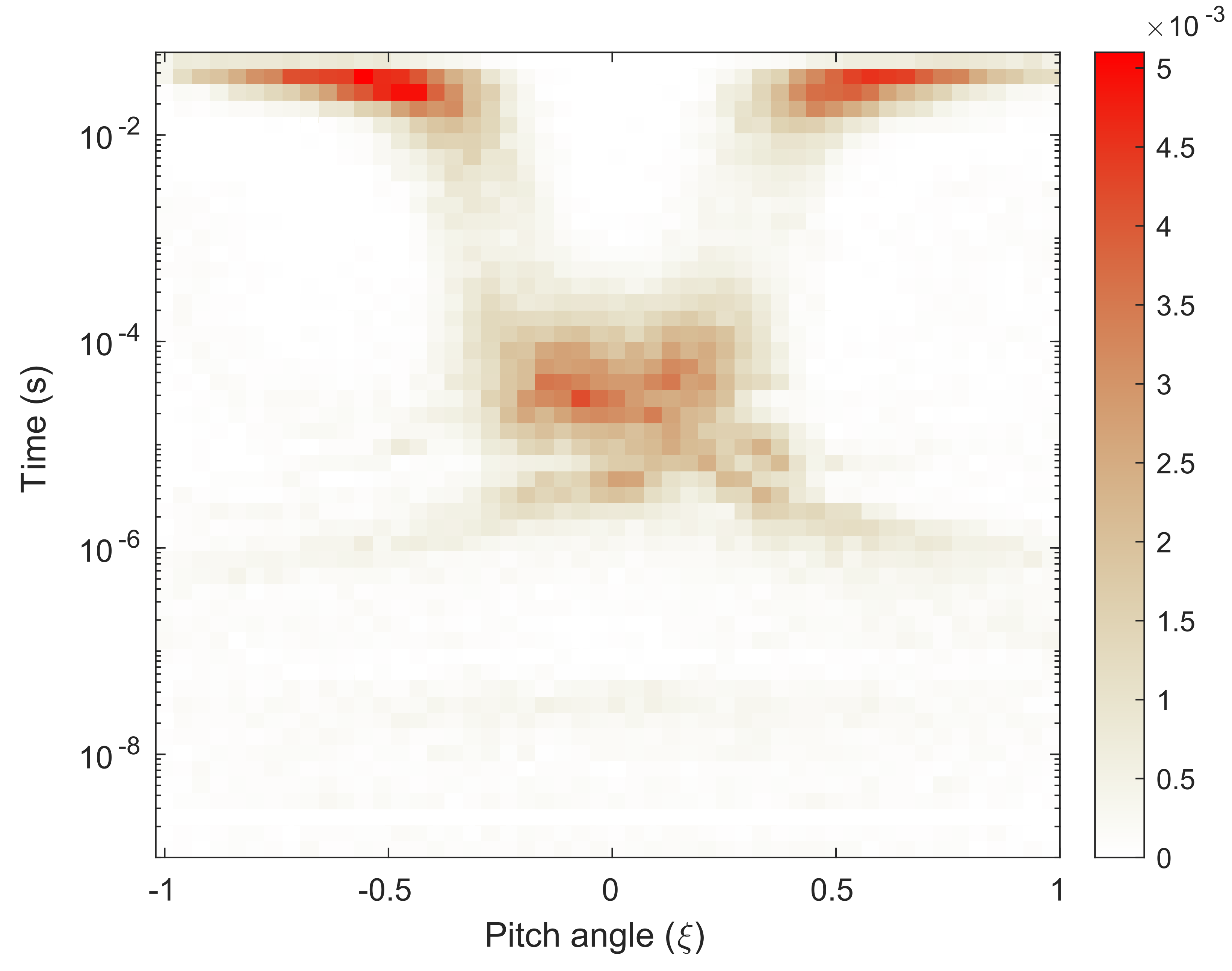}
\caption{Amount of lost particles as a function of initial pitch $\xi$ and loss time for the standard configuration triton slowing-down simulation. The color axis is the normalized number of markers in each histogram slot. The particles can be categorized to first-orbit losses at $t \leq \SI{e-6}{\second}$, losses of trapped orbits at $t \leq \SI{e-2}{\second}$, and collisional losses at $t > \SI{e-2}{\second}$.}
\label{fig:endstate_time_pitch}
\end{figure}

The triton losses are highly dependent on the particle initial pitch, and the triton slowing-down distributions have a distinct gap at low pitch values.
This is due to the fact the energetic particles in this gap are on trapped orbits and are lost from the device via magnetic drifts. They thus have little time to contribute to the fast-ion distribution. This process is illustrated by \cref{fig:endstate_time_pitch}, where the total amount of lost particles is shown as a function of the initial pitch and the time it takes for the particle to be lost.
Less than \SI{2}{\%} of the particles are also lost via first-orbit losses at times less than \SI{e-6}{\second}.
Apart from the collisional losses at $t > \SI{e-2}{\second}$, the tritons are lost at practically their full \SI{1.01}{\mega\electronvolt} energy.

\begin{figure}[tb]
\centering
\includegraphics[width=0.7\textwidth]{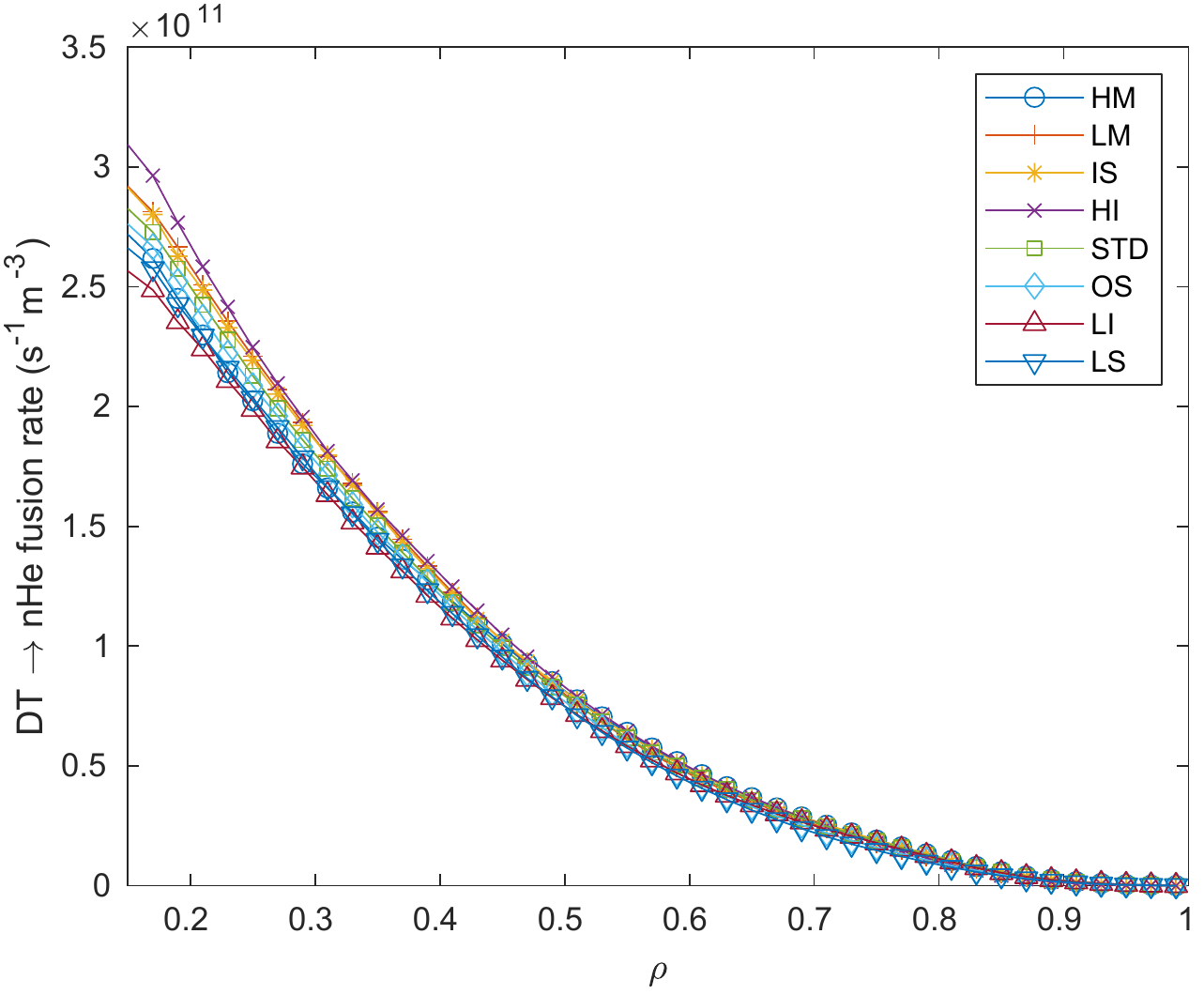}
\caption{Total DT fusion rates as a function of radial coordinate $\rho$ for the different magnetic configurations. The fusion rates are even more centrally peaked and the absolute differences between configurations further reduced compared to the triton birth distributions.}
\label{fig:fusion_profiles_dt}
\end{figure}

An initial AFSI estimation of the DT fusion rate from the reactions between beam deuterium and fast tritons yielded only \SI{e8}{neutrons\per\second}, which is only one per mill of the total neutron birth rate of more than \SI{e11}{neutrons\per\second}.
Due to this, only plasma-triton reactions were included in subsequent analysis.
The reason for the low contribution of the beam-triton reactions is that, from the \SI{1.01}{\mega\electronvolt} triton point of view, the beam ion energy is practically the same as the thermal ion energy and the beam density is more than two orders of magnitude lower than the plasma density.
The radial profiles of the DT fusion rates are shown in \cref{fig:fusion_profiles_dt}.

\begin{figure}[tb]
\centering
\includegraphics[width=0.45\textwidth]{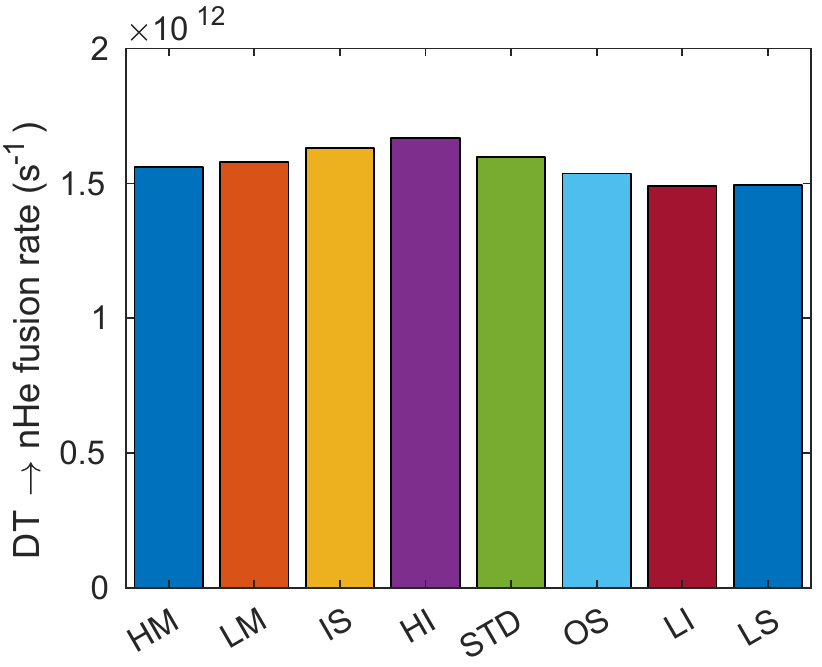}
\caption{Total D-T fusion rates in different magnetic configurations. The relative differences between configurations are smaller than for the D-D fusion rates.}
\label{fig:reaction_contributions_dt}
\end{figure}

The total DT fusion rates in all of the configurations were between \SI{1.49e+12}{\per\second} and \SI{1.67e+12}{\per\second}; the total rates for all configurations are shown in \cref{fig:reaction_contributions_dt}.
Variation between the configurations for the DT fusion rates are smaller than for the DD fusion rates, and the triton burn-up ratio was approximately \SI{0.2}{\percent} in all of the configurations.
Measurements with a SciFi detector at the LHD heliotron have yielded triton burn-up ratios in the same order of magnitude, between \SIrange{0.05}{0.45}{\%} \cite{OgawaBurnup}.
The constant burn-up ratio implies that the DT fusion rate is mainly determined by the triton slowing-down distribution and consequently effected by the triton confinement properties.

\subsection{Sensitivity to kinetic profiles}
\label{sec:results:kinetic}

To assess the sensitivity of the neutron rate on the plasma kinetic profile, one additional simulation was performed for the standard configuration. In this simulation, the central plasma density was halved to \SI{e20}{\per\cubic\meter} and the central temperature doubled to \SI{3}{\kilo\electronvolt}. This maintains the same volume-averaged beta value ($\langle\beta\rangle = \SI{2}{\%}$) while increasing the fusion reactivity and the triton slowing-down time.

The NBI shine-through fraction was \SI{14.4}{\%} with the altered profiles, which is comparable to the reference profiles.
The deuterium power losses, on the other hand, increased from \SIrange{28}{44}{\%}.
This is to be expected due to the longer beam ion slowing-down time; the beam ions have more time to be lost due to drifts before they slow down.
The total \SI{2.45}{\mega\electronvolt} neutron and triton birth rates were found to increase from  \SIrange{6.53e14}{2.45e15}{\per\second} due to the increased beam-plasma and thermonuclear fusion reactivities.
The beam-beam fusion rate also increased by more than an order of magnitude.
The triton power losses however remained virtually identical (\SIrange{47}{48}{\%}).
The total \SI{14.1}{\mega\electronvolt} neutron rate was \SI{1.33e+13}{\per\second}, which is more than eight times larger than for the high-density scenario.
Increasing the plasma temperature thus dominates over the differences between magnetic configurations.

%% file: 05_conclusions.tex
\section{Conclusions and further work}
\label{sec:conclusions}

In this work we have verified that while the confinement of fast ions depends not only on the plasma profiles but also strongly on the magnetic configuration in W7-X, this difference does not extend to the DD and DT fusion rates, which are mainly governed by the kinetic profiles. 
In order to isolate the effect of the magnetic configuration on the neutron rates, the plasma profiles were left unchanged between the configurations.
In reality the temperature and density will differ between magnetic configurations \cite{DinklageConfiguration}.

Since the plasma profiles were kept identical, the only difference in fast ion confinement and the fusion rates comes from the magnetic field configuration.
The magnetic geometry has the largest effect on the fast ion losses near the plasma edge, where most of the NBI ions are born.
Consistently, differences of up to \SI{80}{\%} were found in deuterium beam confinement between configurations, causing significant differences in the slowing-down distribution function.
On the other hand, fusion occurs predominantly in the plasma core, where the configuration effects are weaker.
Due to this, the triton birth rates differ only up to \SI{18}{\%} between configurations.

This configuration difference is smaller for tritons, since they are mainly born in the plasma core; the triton power-loss fraction was between \SIrange{42}{51}{\%} in all of the configurations.
Of the studied magnetic configurations, the high-iota scenario was found to have the highest DT fusion and \SI{14.1}{\mega\electronvolt} neutron production rate, with \SI{1.67e+12}{\per\second} neutrons produced.
Consistently, the low-iota configuration resulted in the least amount of \SI{14.1}{\mega\electronvolt} neutrons, \SI{1.49e+12}{\per\second}.
The differences are nevertheless insignificant compared to the effect of changing the kinetic profiles, where the total \SI{14.1}{\mega\electronvolt} neutron rate was increased by more than \SI{700}{\%} when doubling the plasma temperature while keeping $\left\langle\beta\right\rangle$ constant.

An earlier simple estimation of the neutron propagation to the detector suggests that a total neutron production rate of \mbox{\SI{e12}{\per\second}}, integrated over the plasma volume, would be needed for time-resolved neutron measurements with a SciFi detector \mbox{\cite{KoschinskyTriton}}.
The amount of \mbox{\SI{14.1}{\mega\electronvolt}} neutrons produced in the studied profiles and configurations exceeds this limit.
However, the total NBI power of \mbox{\SI{15.1}{\mega\watt}} is not expected to be available while simultaneously operating the ECRH system.
A practically feasible amount of NBI power would be \mbox{\SI{7.84}{\mega\watt}}, i.e., using half of the injectors.
This adjustment would approximately halve the beam-target and beam-beam fusion rates presented in this work, depending on which PINIs are operated, limiting time-resolved SciFi measurements to high-performance phases with higher ion temperature.

In this work a thorough scan of profiles shape and collisionality was excluded due to the time-consuming simulation method used.
Earlier ASCOT simulations with hydrogen beams showed that the main source of differences in beam-ion confinement is the density profile \mbox{\cite{AkaslompoloW7-XWallLoads}}.
A comprehensive temperature and density scan for neutron production in W7-X is better suited for a simple 1D model \mbox{\cite{KoschinskyTriton}}.

A more reliable estimate of the neutron signal requires simulating the \mbox{\SI{14.1}{\mega\electronvolt}} neutron propagation to the detector.
This can be done with a Monte Carlo neutron transport code -- such as Serpent \mbox{\cite{LeppanenSerpent}} -- and requires a full 3D representation of W7-X device and its materials.
The Serpent model of W7-X has recently been completed, and a proof-of-concept calculation produced \mbox{\SI{158}{counts\per\second}} from triton burnup neutrons in the standard configuration case presented in this work.
This results in an estimated SciFi integration time of \mbox{\SI{50}{\milli\second}} \mbox{\cite{Äkäslompolo2020serpent}}.
In any case, a realistic estimate of the neutron counts to the detector can only be made after experimental plasma profiles from the last W7-X campaign become available.

Further analysis of triton orbits and especially triton losses in W7-X with ASCOT could be beneficial in calculating triton losses to the W7-X wall.
Determining triton loss patterns is important not only for heat load management but also for radiation safety: due to their high energy the tritons produced in DD can be deposited deeply into the device walls, which leads to tritium retention over time \cite{Kurki-SuonioTriton}.
Studying triton orbits is also beneficial for designing future stellarator reactors, because the gyroradius $r_L$ of \SI{1.01}{\mega\electronvolt} tritons in W7-X is similar to \SI{3.6}{\mega\electronvolt} alpha particles in a foreseen HELIAS reactor.
These particles are nevertheless not directly analogous because the normalized gyroradius $r_* = r_L / a$, where $a$ is the machine minor radius, of particles would be smaller in a HELIAS due to the larger machine size and higher magnetic field.

%% file: 06_acknowledgments.tex
\ack

The calculations were performed on Marconi-Fusion, the High Performance Computer at the CINECA headquarters in Bologna (Italy). The computational resources provided by Aalto Science-IT project are also acknowledged. This work was partially funded by the Academy of Finland project No. 298126. This work has been carried out within the framework of the EUROfusion Consortium and has received funding from the Euratom research and training programme 2014-2018 and 2019-2020 under grant agreement No 633053. The views and opinions expressed herein do not necessarily reflect those of the European Commission.